\documentclass{ifacconf}

\usepackage{graphicx}      
\usepackage{natbib}        
\usepackage{enumerate}

\usepackage{amsmath} 
\usepackage{amssymb} 
\usepackage{amsfonts}
\usepackage[usenames,dvipsnames]{xcolor}

\def\Neil#1{{{\color{red}\texttt{ #1}}}}

\newcommand{\R}{\mathbb{R}}

\newcommand{\argmin}{\operatornamewithlimits{argmin}}

\newcommand{\twopartdef}[4]
{
	\left\{
		\begin{array}{ll}
			#1 & \mbox{if } #2 \\
			#3 & \mbox{if } #4
		\end{array}
	\right.
}

\newcommand{\PGS}[1]{\normalsize{{\color{Red} \ #1}}}
\newcommand{\maryam}[1]{\normalsize{{\color{magenta}MK:\ #1}}}

\newtheorem{theorem}{Theorem} 
\newtheorem{lemma}{Lemma}  
\newtheorem{definition}{Definition} 
\newtheorem{example}{Example} 

\begin{document}
\begin{frontmatter}

\title{Exploring the Vickrey-Clarke-Groves Mechanism for Electricity  Markets\thanksref{footnoteinfo}} 

\thanks[footnoteinfo]{This work is partially funded under M. Kamgarpour's European Union ERC Starting Grant CONENE.}

\author[First]{Pier Giuseppe Sessa} 
\author[Second]{Neil Walton} 
\author[Third]{Maryam Kamgarpour}

\address[First]{Automatic Control Laboratory, D-ITET, ETH Z\"{u}rich\\ (e-mail: sessap@student.ethz.ch).}
\address[Second]{School of Mathematics, University of Manchester\\ (e-mail: neil.walton@manchester.ac.uk)}
\address[Third]{Automatic Control Laboratory, D-ITET, ETH Z\"{u}rich\\ (e-mail: mkamgar@control.ee.ethz.ch)}

\begin{abstract}                
Control reserves are power generation or consumption entities that ensure balance of supply and demand of electricity in real-time. In many countries, they are operated through a market mechanism in which entities provide bids. The system operator determines the accepted bids based on an optimization algorithm.  We develop the Vickrey-Clarke-Groves (VCG) mechanism for these electricity markets. We show that all advantages of the VCG mechanism including incentive compatibility of the equilibria and efficiency of the outcome can be guaranteed in these markets. Furthermore, we derive conditions to ensure  collusion and shill bidding are not profitable. Our results are verified with numerical examples.
\end{abstract}

\begin{keyword}
Electrical networks, game theory, optimization, control reserves.
\end{keyword}

\end{frontmatter}

\section{Introduction}

The liberalization of electricity markets leads to opportunities and challenges for ensuring stability and efficiency of the power grid. For a stable grid, the supply and demand of electricity at all times need to be balanced. This instantaneous balance is reflected in the grid frequency. Whereas scheduling (yearly, day-ahead)  is based on forecast supply and demand of power, the \textit{control reserves} (also referred to as ancillary services) provide additional controllability to balance supply and demand of power  in  real-time.  With increasing volatile renewable sources of energy, the need for control reserves also has increased. This motivates  analysis and design of  optimization algorithms and market mechanisms that procure these reserves. 

The objective of this paper is a game theoretic exploration of an alternative market mechanism for the control reserves with potential  improvements. To  further discuss this, we briefly discuss relevant features of the existing market mechanism.  Control reserves are categorized as primary, secondary, and tertiary. Primary reserves balance frequency deviations in timescale of seconds.  Secondary reserves balance the deviations on a timescale of seconds to minutes not resolved by primary control. Tertiary reserves restore secondary reserves and typically act  15 minutes after a disturbance to frequency. The secondary and tertiary control reserves in several countries are procured in a market. In the Swiss market for example, the auction mechanism implemented by the Transmission System Operator (TSO) minimizes the cost of procurement of required amounts of power, given bids \citep{abbaspourtorbati2016swiss}.

In a pay-as-bid mechanism, since payments to winners are equal to their bid prices, a rational player may over-bid to ensure  profit. As an alternative to pay-as-bid, we explore the \textit{Vickrey Clarke Groves} (VCG)  mechanism. This is one of the most prominent auction mechanisms. The first analysis of the VCG mechanism was carried out by \citep{vickrey1961counterspeculation} for the sale of a single item. This work was subsequently generalized to multiple items by \citep{clarke1971multipart} and \citep{groves1973incentives}. 
  
It has been shown that the VCG mechanism is the only mechanism that possesses \emph{efficiency} and \emph{incentive compatibility}. Efficiency implies that goods are exchanged between buyers and sellers in a way that creates maximal social value. Incentive compatibility means that it is optimal for each participant to bid their \emph{true value}. Variants of the VCG mechanism have been successfully deployed generating billions of dollars in Spectrum auctions, for instance, in the 2012 UK spectrum auction \citep{cramton2013spectrum,day2012quadratic} and in advertising, for instance, by Facebook\footnote{https://developers.facebook.com/docs/marketing-api/pacing}  \citep{varian2014vcg}. For further discussion on the VCG mechanism and its application to real auctions we recommend \citep{milgrom2004putting,klemperer2004auctions}.

Investigation must be performed before applying the VCG mechanism. As outlined in the paper of Ausubel and Milgrom \citep{ausubel2006lovely}, coalitions of participants can influence the auction in order to obtain higher collective profit. These peculiarities occur when the outcome of the auction is not in the \textit{core}. The core is a solution concept in coalition game theory where prices are distributed so that there is no incentive for participants to leave the coalition \citep{osborne1994course}. This has recently motivated the study and application of VCG auctions where the outcome is projected to the core \citep{cramton2013spectrum, abhishek2012incentive}.

The electricity market can be thought of as a reverse auction. In contrast to an auction with multiple goods, in an electricity market, each participant can bid for continuum values of power. Furthermore, to clear this market, certain constraints, such as balance of supply and demand and network constraints need to be guaranteed. Due to the differences between an electricity market and an auction mechanism for multiple items (such as spectrum or adverts), there are conceptual and theoretical advances in VCG mechanism that need to be analyzed. 

In this paper, we apply the VCG mechanism to control reserve markets and provide a mathematically rigorous analysis of it.  We show that efficiency and incentive compatibility of the VCG mechanism will hold even in the case of stochastic markets, see Theorem \ref{thm:incentive_comp}. On the other hand, we provide examples where shill bidding might occur. The remainder of the paper develops ways to resolve this issue. In particular, building upon a series of results based on coalitional game theory, in Theorem \ref{thm:no_collusions} we show how a simple pay-off monotonicity condition removes incentives for shill bidding and other collusions. The proofs developed significantly simplify the arguments of Ausubel and Milgrom \citep{ausubel2006lovely}.

The paper is organized as follows. In Section \ref{sec:setup} we introduce the VCG mechanism for control reserve markets, analyzing its positive and negative aspects. Throughout Section \ref{sec:solution_approach} we investigate conditions that can mitigate these problems making the mechanism competitive. We conclude with specific simulations based on data available from Swissgrid (the Swiss TSO) showing the applicability of VCG mechanism to the Swiss ancillary service market.


\section{Electricity auction market setup} \label{sec:setup}
We briefly describe the control reserve market of Switzerland. The formulation and results derived are generalizable to alternative markets, with similar features as will be discussed. The Swiss system operator (TSO), Swissgrid, procures secondary and tertiary reserves in its reserves markets. These consist of a weekly market where secondary reserves are procured and daily markets where both secondary and tertiary reserves are procured. Each market participant submits a bid that consists of a price  per unit of power (CHF/MW, swiss franc per megawatt) and a volume of power which it can supply (MW).  Offers are indivisible and thus, must be accepted entirely or rejected.  Moreover, \emph{conditional offers} are accepted. This means that a participant can offer a set of bids, of which only one can be accepted.  If an offer is accepted, the participant is paid for its availability irrespective of whether these reserves are deployed (an additional payment is made in case of deployment). This availability payment, under the current swiss reserve market, is pay-as-bid. An extensive description of the Swiss Ancillary  market is given in \citep{abbaspourtorbati2016swiss}.

We abstract the control reserve market summarized above as follows. Let $L$ denote the set of auction participants and $| L | = N$. Let $B_j = (c_j,p_j)$ be all the bids placed by participant $j$, where $p_j \in \R^{n_j}$ is the vector of power supplies offered (MW) and $c_j \in \R^{n_j}$ are their corresponding requested costs (or prices). Here $n_j$ is the number of bids from participant $j$. Let $B = \{B_j, j\in L\}$ be the set of all bids and $n = \sum_{j=1}^n n_j$. Given a set $B$, a  mechanism  defines which bids are accepted with a \emph{choice function}, $f(B) \in \{0,1\}^n$ and a payment to each participant, \emph{payment rule} $q_j(B)$. The \emph{utility} of participant $j$  is hence  
\begin{equation}\label{def:utilities}
u_j(B) = q_j(B) - \bar{c}_j^\top f_j(B),
\end{equation}
where $\bar{c}_j\in \R^{n_j}$ is participant $j$'s true cost of providing the offered power $p_j$ and $f_j(B) \in \{0, 1\}^{n_j}$ is the binary vector indicating his accepted bids.

The transmission system operator's objective function is
\[
J(x,y;B) = {c}^\top x + D(x,y).
\]
The variable $x \in \{0,1\}^n$ selects the accepted bids, $y \in \R^p$ can be any additional variables entering the TSO's optimization and $D: \{0,1\}^n \times \R^p \rightarrow \R$ is a general function.
In most electricity market, the objective is to minimize the cost of procurement subject to some constraints: 
\begin{subequations}\label{eq:general_clearing_model}
\begin{align}
& J^\star(B) = \hspace{0.5em}  \min_{x , y } \hspace{0.5em} J(x,y;B) \quad \text{s.t.} \hspace{1em} g(x,y, {p}) \leq 0\\
& x^\star(B) = \hspace{0.5em}  \argmin_{x }  \Big\{ \min_{y: g(x,y, {p}) \leq 0} J(x,y;B) \Big\}
\end{align}
\end{subequations}
The above constraints correspond to procurement of the required amounts of power,  e.g. in the Swiss reserve markets accepted reserves must have a deficit probability of less than 0.2\%. 
We let $X$ be the feasible values of $x$ for this optimization. 
The optimization defines a general class of  models, where the cost function is affine in ${c}$ and the prices of bids do not enter the constraints. 
\subsection{The pay-as-bid  mechanism}

In the current pay-as-bid mechanism we recognize:
\begin{align*} 
&f(B) = x^\star(B) \\ & q_j(B)=c_j^\top x_j^\star(B), \hspace{1em}  j \in L.
\end{align*}
It follows that each participant's utility is $u_j(B)= (c_j - \bar{c}_j)^\top x_j^\star(B)$. As such, rational participants would bid more than their true values to make profit. Consequently, under pay-as-bid, the TSO attempts to minimize inflated bids rather than true costs. Thus,  pay-as-bid cannot guarantee power reserves are procured cost effectively.


\subsection{The VCG  mechanism}
The VCG mechanism is characterized with the same choice function as the pay-as-bid mechanism but a different payment rule. 
\begin{definition}
\label{def:VCG}
The \emph{Vickrey-Clarke-Groves} (VCG) choice function and payment rule are defined as:
\begin{align*} & f(B) = \argmin_{x \in X}J(x,y;B)=x^\star(B), \\ & q_j(B)= h(B^{-j}) - \big(J^\star(B) - c_j^\top x_j^\star(B)\big) \hspace{1em} \forall j \in L,
\end{align*} 
where $B^{-j}$ denotes the vector of bids placed by all participants excluding $j$. The function $h$ must be carefully chosen to make the mechanism meaningful. Namely, we require that payments go from the TSO to power plants, \emph{positive transfers}, and that power plants will not face negative utilities participating to such  auctions, \emph{individual rationality}. A particular choice of $h$ is the \emph{Clarke pivot-rule}, which minimizes  the procurement cost given all bids excluding $j$'s: $$h(B^{-j})= J^\star(B^{-j}).$$ 
\end{definition}
	
A set of bids $B = \{B_j, j\in L\}$ is a \emph{dominant-strategy Nash equilibrium} if for each  participant $j$,
		$$ \hspace{1em}  u_j(B_j,B^{-j}) \geq u_j(\tilde{B}_j, B^{-j}) \hspace{1em} \forall \tilde{B}_j,  \forall B^{-j}. $$
 Moreover, a dominant-strategy equilibrium is \emph{incentive compatible} if $B_j=(\bar{c}_j,p_j)$ where $\bar{c}_j$ is the true cost of power $p_j$, as given in \eqref{def:utilities}. That is, each participant finds it more profitable to bid truthfully $B_j$, rather than any other vector $ \tilde{B}_j$, regardless of other participants' bids. Hence, all the bidding strategies are \emph{dominated} by strategy $B_j$. 

The following theorem summarizes the contributions of  \citep{vickrey1961counterspeculation}, \citep{clarke1971multipart} and \citep{groves1973incentives} in designing the VCG mechanism. In our proof, we are mindful of the slightly non-standard setting of the electrical markets: that auctions are ``reverse-auctions", i.e. with a single buyer and many sellers, and that constraints in the optimization problem may be non-standard.

\begin{theorem}
\label{thm:incentive_comp}
Given the clearing model of \eqref{eq:general_clearing_model}. 
\begin{enumerate}[a)]
\item The energy procurement auction under VCG choice function and payment rule is a \emph{Dominant-Strategy Incentive-Compatible} (D.S.I.C) mechanism. 
\item The VCG outcomes are \emph{efficient}, that is, the sum of all the utilities is maximized. 
\item The Clarke pivot rule ensures {positive transfers} and {individual rationality}.
\end{enumerate}
\end{theorem}

\begin{pf}a) We distinguish between the participant $j$ placing a generic bid $\tilde{B_j} = (c_j,p_j)$ and biding truthfully $\bar{B_j}= (\bar{c_j},p_j)$.  For $ \tilde{B}=(\tilde{B_j},B^{-j})$, substituting the VCG choice function and payment rule with $J^\star(\tilde{B})$ as in \eqref{eq:general_clearing_model}:
\vspace{-0.3em}
\begin{align*}
&u_j(\tilde{B})= h(B^{-j}) - \\ 
&\big(\sum_{i\neq j}{c_{i}^\top x_{i}^{\star}(\tilde{B})} + {\bar{c}_j}^\top x_j^{\star}(\tilde{B}) + D(x^{\star}(\tilde{B}), y^{\star}(\tilde{B})) \big),
\end{align*} 
where the term in brackets is the cost $J$ of $(x^\star(\tilde{B}),y^{\star}(\tilde{B}))$ but evaluated at $(\bar{B_j}, B^{-j})$. For $\bar{B} = (\bar{B_j}, B^{-j})$, however, $ u_j(\bar{B})= h(B^{-j}) - J^\star(\bar{B})$. Note that
\begin{align*}
J^*(\bar{B}) \leq \big(\sum_{i\neq j}{c_{i}^\top x_{i}^{\star}(\tilde{B})} + {\bar{c}_j}^\top x_j^{\star}(\tilde{B}) + D(x^{\star}(\tilde{B}), y^{\star}
\end{align*}
We then have $u_j(\bar{B}) \geq u_j(\tilde{B})$ because $(x^\star(\tilde{B}),y^{\star}(\tilde{B}))$ is a feasible suboptimal allocation for the available bids $\bar{B}$.

b) Let $u_0(B)$ denote the utility gained by the TSO, that is, $u_0(B)= - \big( \sum_{j=1}^{N}{q_j(B) + D(x^\star(B), y^\star(B))\big)}$.
By Definition \eqref{def:utilities} and incentive compatibility, $q_j(B)= u_j(B) + c_j^\top x_j^\star(B)$. We then have: $u_0(B) = - J^\star(B) - \sum_{j=1}^{N}{u_j(B)}$. Hence, $  \sum_{j=0}^{N}{u_j(B)} = - J^\star(B)$, which is maximized by the clearing model \eqref{eq:general_clearing_model}.

c) This can be easily verified substituting $h(B^{-j})$:
\begin{align}
q_j(B) &= J^\star(B^{-j}) - \big(J^\star(B) - c_j^\top x_j^\star(B)\big)\notag \\
& = c_j^\top x_j^\star(B) + \big( J^\star(B^{-j}) - J^\star(B)\big) \geq 0 \hspace{2em} \forall B,\notag \\
u_j(B) &= J^\star(B^{-j}) - J^\star(B) \geq 0 \hspace{2em} \forall B.\label{eq:util}
\end{align}
\vspace*{-.3cm}
\qed
\end{pf}
In summary, all producers have  incentive to reveal their true values for price of power in a VCG market. Thus, it becomes easier for entities to enter the auction, without spending resources in computing optimal bidding strategies. This can help in achieving market  liberalization objectives. Moreover, from the above theorem it follows that the winners of the auctions are   the producers with the lowest true values. This is because participants bid truthfully and the VCG choice function minimizes the cost of the accepted bids. 

Note that the result above are very general. We do not need to assume any particular form for the term $D$ and the constraints $g$. Furthermore, using the  exact same approach in the proof, we can state the exact same theorem for the case in which the participants provide continuous bid curves: $c_i : P_i \rightarrow \R_+$, where $P_i \subset \R_+$. The difference is only in notation; we use $p_i^\star$ for the optimal bid accepted from player $i$ instead of the binary variables $x^\star$. 

So, there are persuasive arguments for considering VCG market for control reserves. However, there are potential disadvantages that must be eliminated. 
\begin{example}
\label{example_1}
Suppose the TSO has to procure 800 MW from PowerPlant1, $PP_1$, who bids $40'000$ CHF for 800 MW, and PowerPlant2, $PP_2$, who bids $50'000$ CHF for 800 MW. Under the VCG mechanism, PowerPlant1 wins the auction receiving a payment of $50'000$ CHF. Suppose now that power plants $PP_3, PP_4, PP_5$ and $PP_6$ entered the auction each bidding 0 CHF for 200 MW. Clearly, the new entrants become winners and each of them would receive a VCG payment of $40'000$ CHF.

This example shows that: (a) producers with very low prices (in this case 0 CHF) could receive very high payments; (b) \emph{collusion} or \emph{shill bidding} can increase participants' profits. In fact, $PP_3, PP_4, PP_5$ and $PP_6$ could be a group of losers who jointly lowered their bids to win the auction, or they could represent multiple identities of the same losing participant (i.e. a power plant with true value greater than 40'000 CHF for 800 MW). Entering the auction with four shills, however, this participant would have received a payment of 4$\times$40'000 CHF. 
\end{example}
Our goal is now to derive conditions that make VCG outcomes competitive and prevent shill bidding or collusion. 

\section{Solution approach for VCG market}
\label{sec:solution_approach}
In coalition game theory, the \emph{core} is the set of allocations of goods that cannot be improved upon by the formation of coalitions. \citep{ausubel2006lovely} identify conditions for a VCG outcome to lie in the core. Following their analysis we derive conditions for core outcomes in our setting and provide new simpler proofs relevant to our problem formulation that show that shill bidding and collusion can be eliminated from certain class of electricity  markets under the VCG mechanism.

 Given a game where $L$ is the set of participants, let  $w$ denote the coalitional value function
\begin{align*}
w(S) = \twopartdef { - J^\star(S) }{0 \in S \subseteq L}{0} { 0 \notin S\subseteq L}.
\end{align*} 
This function provides the optimal objective function, for any subset of participants $S$ that includes the TSO.
Here, $J^\star(S)$ is the cost  the TSO incurs for the VCG outcome with participants $S$. That is, $J^\star(S)$ is the solution to optimization \eqref{eq:general_clearing_model} with $c_j=\bar{c}_j$ for all $j$, and with additional constraints that $x_j=0_{n_j\times 1}$ for all $j\notin S$. Clearly $J^\star(S) \leq J^\star(S')$ for $S'\subset S$ since increasing participation reduces costs.
We thus let $(L,w)$ represent the coalition game associated with the auction.
\begin{definition}
The $Core(L,w)$ is defined as follows
\begin{align*}
\bigg\{ u \in \mathbb{R}^{n+1}
	\mid \sum_{j=0}^{N}u_j=-J^\star(L) ,\hspace{0.2em} w(S) \leq \sum_{j \in S}u_j \hspace{0.2em}\forall S \subseteq L \bigg\}.
\end{align*}
The core is thus the set of all the feasible outcomes, coming from an \emph{efficient} mechanism (first equality above), that are \emph{unblocked} by any coalition (the inequality).
We say that an outcome is competitive if it lies in the core; that is, there is no incentive for forming coalitions. In the previous example, the outcome was not competitive because it was blocked by coalition $\{0,1\}$.  PowerPlant1 was offering only 40'000 CHF for the total amount of 800 MW. It will be also shown in Theorem \ref{thm:no_collusions} that core outcomes eliminate any incentives for \emph{collusions} and \emph{shill bidding}.
\end{definition}
\subsection{Ensuring core payments}
Since core outcome is a competitive outcome, we investigate under which conditions the outcomes of the VCG mechanism applied to the control reserve market will be in the core. Note that there are $2^L$ constraints that define a core outcome. Our first Lemma provides an equivalent characterization of  the  core  with significantly lower number of constraints. 
\begin{lemma}
\label{lem:lemma_core}Given a VCG auction $(L,w)$, let $u=[u_0 \dots u_N]$ be its outcome and $W\subseteq L$ the corresponding winners. Assuming participants revealed their true values, $[u_0 \dots u_N] \in Core(L,w)$ if and only if, $\forall K\subseteq W$,
\begin{equation}
\label{eq:core_constraints}
 \sum_{j \in K }{\big( J^\star(L^{-j}) - J^\star(L) \big)} \leq J^\star(L \setminus K) - J^\star(L).
\end{equation}
\end{lemma}

\begin{pf} Core constraints with $0\notin S$ are immediately satisfied as $u_j\geq 0$ (individual rationality, Theorem 1c). Now,
$u$ is unblocked by any $0 \in S\subseteq L \iff -J^\star(S) \leq \sum_{j \in S\setminus 0}u_j + u_0 = \sum_{j \in S\setminus 0}u_j - \sum_{j \in W} u_j - J^\star(L)$ (since $u_0 = - J^\star(L) - \sum_{j\in W}{u_j}$ is the outcome with $L$ participants). Thus, $\sum_{j \in W\setminus S}u_j \leq J^\star(S)- J^\star(L)$. Moreover, fixing $K = W\setminus S$, the dominant constraints are those corresponding to minimal $J^\star(S)$, in particular, when $S = L\setminus K$ (this being maximal set with $K$ not taking part in the coalition $S$).
Finally recall from \eqref{eq:util} that, under the VCG mechanism, $u_j = J^\star(L^{-j}) - J^\star(L)$. \qed
\end{pf}
The following definition and theorem act over subsets of participants. Here, we imagine that there is a set of potential participants $Z$ and, for each subset $L$ of $Z$, we consider whether the outcome of the auction with L participants lies in the core.  
\begin{definition}
Participant $j\in Z$ displays \emph{payoff monotonicity} if $\forall 0 \in S \subseteq S' \subseteq Z$,
\begin{equation}
\label{eq:payoff_monot}
u_j(S')\! =\! J^\star(S'^{-j})\! -\! J^\star(S')\!\leq\!  J^\star(S^{-j})\! -\! J^\star(S)\!=\! u_j(S) 
\end{equation}
\end{definition}

\begin{theorem}
\label{thm:payoff_mon}
The outcome of the VCG auction~$(L,w)$ lies in the core for all $L\subseteq Z$ if and only if \emph{payoff monotonicity} holds for each participant in $Z$.
\end{theorem}

%

\begin{pf}
To prove that payoff monotonicity is sufficient for $u$ to lie in the core, we prove that \eqref{eq:core_constraints} holds. Let $K=\{ j_1 \dots j_k\}$. Considering $\sum_{j \in K}{\big(J^\star(L^{-j}) - J^\star(L)\big)}$, we notice that: 
$ J^\star(L^{-j_1}) - J^\star(L) \leq J^\star(L\setminus K) - J^\star(L^{-\{j_2, \dots, j_k \}})$ since $j_1$ displays payoff monotonicity over $S=L^{-\{j_2, \dots, j_k \}} \subseteq  L= S'$; we also have $J^\star(L^{-j_2}) - J^\star(L) \leq J^\star(L^{-\{j_2, \dots, j_k \}}) - J^\star(L^{-\{j_3, \dots, j_k \}})$ since $j_2$ displays payoff monotonicity over $S = L^{-\{j_3, \dots, j_k \}} \subseteq L = S'$. We can continue with the same considerations up to $J^\star(L^{-j_{k-1}}) - J^\star(L) \leq J^\star(L^{-\{j_{k-1}, j_k \}}) - J^\star(L^{-j_k})$ since $j_{k-1}$ displays payoff monotonicity over $S=L^{-j_k} \subseteq L = S'$. Therefore, $\sum_{j \in K } J^\star(L^{-j}) - J^\star(L) \leq J^\star(L\setminus K) - J^\star(L^{-\{j_2, \dots, j_k \}}) + J^\star(L^{-\{j_2, \dots, j_k \}}) + \dots - J^\star(L) = J^\star(L \setminus K) - J^\star(L) $. This same argument holds for any subset of participants $L$ and $\forall K\subseteq W$. Thus \eqref{eq:core_constraints} holds and so, by Lemma \ref{lem:lemma_core}, the VCG outcome belongs to the core. \\
To prove that payoff monotonicity is also necessary for outcomes to lie in the core, suppose that ${j}$ does not display payoff monotonicity. Then, there exist sets ${S}$,${S'}$ where \eqref{eq:payoff_monot} does not hold. We may chose $S'= S\cup \{i\}$ for some $i$. To see this, take $S=S_0$ and $S_k=S\cup \{ j_k\}$ with $S_\kappa=S'$, then, since payoff monotonicity does not hold,
\begin{align}
&\sum_{k=1}^\kappa J^\star(S^{-j}_k)-J^\star(S^{-j}_{k-1}) = J^\star(S'^{-j})-J^\star(S^{-j}) \label{interpolate}\\
&\qquad >  J^\star(S')-J^\star(S) = \sum_{k=1}^\kappa J^\star(S_k)-J^\star(S_{k-1}).\notag
\end{align}
The strict inequality above must hold for one of the summands  $J^\star(S^{-j}_k)-J^\star(S^{-j}_{k-1}) > J^\star(S_k)-J^\star(S_{k-1})$.  So we may consider sets ${S} \subseteq {S'}$ that differ by one participant, say $i$. Let $S'$ and $S$ be the minimal such sets.
By minimality,
$ u_j(S')=J^\star({S'}^{-{j}}) - J^\star({S'}) > J^\star({S}^{-{j}}) - J^\star({S}) \geq J^\star({S'}^{-i-{j}}) - J^\star({S'}^{-i})\geq 0$ for $i \in {S'}\setminus {S}$.
Further, after rearranging the above inequality we see that $u_i(S')= J^\star(S'^{-i})-J^\star(S') > J^\star(S'^{-i-j})-J^\star(S'^{-j})\geq 0$.
That is \emph{both} participant $i$ and $j$ are winners of the VCG auction with participants $S'$ (instead of $L$).  
Then, considering Lemma \ref{lem:lemma_core} for the auction $({S'},w)$ with  $i,{j}\in W$, and ${K}=\{i, {j}\}$, we have: $\sum_{j' \in {K} }{ J^\star({S'}^{-j'}) - J^\star({S'})} > J^\star({S'}^{-{j}-i}) - J^\star({S'}^{-i}) + J^\star({S'}^{-i}) - J^\star({S'}) = J^\star({S'} \setminus {K}) - J^\star({S'})$.
Thus \eqref{eq:core_constraints} does not hold since the outcome $u(S')$ is blocked by coalition $S'\setminus {K}$.\qed
\end{pf}
Whether a VCG outcome is competitive hence depends  on a particular property of the optimal cost $J^\star$. Namely, $J^\star$  has to make \eqref{eq:payoff_monot} hold for each $j$. Note that a similar result was proven in  \citep{ausubel2006lovely}, for a sale auction of a finite number of objects, without any constraints. Our result generalizes this to markets with continuous  goods and arbitrary social planner objectives of the form \eqref{eq:general_clearing_model}.

\subsection{Single stage electricity procurement auction}
The class of auctions cleared by \eqref{eq:general_clearing_model} is very general and suitable for mechanisms with multiple stages of decisions. We will see, in fact, how the two-stages Swiss clearing model described in \citep{abbaspourtorbati2016swiss} can be abstracted as in \eqref{eq:general_clearing_model}. But first, we start considering simpler auctions, characterized by single-stage decisions. More specifically, energy procurement auctions where the TSO has to procure a fixed amount of $M$ MW, subject to conditional offer constraints. Hence, we consider auctions cleared by:
\begin{subequations}\label{clearing}
\begin{align} 
\label{eq:simpler_clearing_model}J^\star(S) =&  \min_{x} {\sum_{j \in S}{c_j^\top x_j}} \\
\text{s.t.} \quad &   \sum_{j \in S}{p_j^\top x_j} \geq M, \nonumber \\ 
&  1_{n_j\times 1}^\top x_j \leq 1 \hspace{1em} \forall j \in S \nonumber
\end{align}
Note that the last constraint above ensures that each bidder can only have one offer accepted. We further suppose that the power offered by participants is equally spaced by some increment $m$, which is a divisor of $M$ and is chosen by the TSO: 
\begin{equation}\label{bids}
\text{each $j$ bids on power offers $p_j^{(k)}=k m$, ${k\in\mathbb Z}$.}
\end{equation}
\end{subequations}

The model above is a simple clearing model within  class \eqref{eq:general_clearing_model}. We can now derive conditions on participants' bids to ensure pay-off monotonicity, condition \eqref{eq:payoff_monot},  is satisfied. Thus, we derive conditions under which the outcome of  auctions cleared by \eqref{eq:simpler_clearing_model}, \eqref{bids} would lie in the core.

\begin{figure}[h]
	\begin{center}
		\includegraphics[scale=0.6]{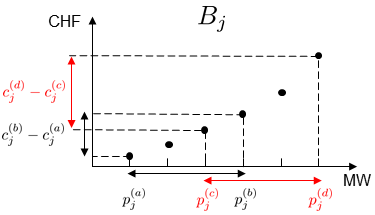}
		\caption{Bids in $B_j$ satisfying condition \eqref{(i)}.}
		\label{fig:condition_bids}
	\end{center}
\end{figure}

\begin{theorem}
\label{thm:conditions_on_bids}
Given \eqref{eq:simpler_clearing_model}, \eqref{bids} if 
$p_j^{(b)} - p_j^{(a)} = p_j^{(d)} - p_j^{(c)}> 0$ with  $0 \leq p_j^{(a)} <p_j^{(c)}$ implies that 
\begin{equation}\label{(i)}
c_j^{(d)} - c_j^{(c)} > c_j^{(b)} - c_j^{(a)}
\end{equation}
for each $j \in Z$,
then bidders satisfy payoff monotonicity condition \eqref{eq:payoff_monot} under the VCG payment rule.
\end{theorem}
In words, marginally increasing cost condition \eqref{(i)} implies core outcomes, and thus eliminates incentives for collusions. Condition \eqref{(i)} is visualized in Figure \ref{fig:condition_bids}.

To prove that condition \eqref{(i)} is sufficient for payoff monotonicity, the following Lemma is needed. 
\begin{lemma}\label{lemma_regularity}
Under clearing model \eqref{clearing}, for an auction with participants $S$ and $S'=S\cup\{i\}$  with corresponding optimal power allocations $p$ and $p'$,  condition \eqref{(i)} implies that 
	\[\forall j\in S, \quad
	p_j'\leq p_j.
	\]
\end{lemma}
In the following proofs we apply the notation that if $p_j$ is the accepted power allocation from  bidder $j\in S$, then $c_j^{(p_j)}$ is the associated cost bid from $j$. If the accepted allocation is zero, we define $c_j^{(0)} = 0$. 
\begin{pf}
 The proof  follows by contradiction. That is, we will show that when $p'$ is such that ${p}'_{{j}} > p_{{j}}$, for some $j\in S$ then $p'$ can be modified to provide a lower cost allocation, $q$, for participants  $S$ (thus contradicting optimality of $p$). First, we notice that since bids are equally spaced by $m$ MW \eqref{bids} and satisfy condition \eqref{(i)}, it is never optimal to accept more than $M$ MW ($\sum_{j \in S}p_j = \sum_{j \in S'}{p'_j} = M$).

Now, assume $p'$  is such that ${p}'_{{j}} > p_{{j}}$, for some $j\in S$. In order to procure exactly $M$ MW, some participants' accepted MWs must decrease, that is, the set $K = \{ j \in S \, | \, p_j'< p_j  \}$ is non-empty. Consider a feasible allocation $q'$ for the auction with $S'$ participants where $M$ units of power are procured and
\[
q'_j = 
\begin{cases}
p'_i, &\text{ for }j=i\\
p_j, & \text{ for } j\in S\backslash K \\
q'_j &\text{ for }j\in K \text{ where }p_j' \leq q'_j \leq p_j
\end{cases}
\]
So, $q'$ is constructed from $p'$ by transferring $m'=\sum_{i\in S\backslash K} p'_j - p_j $ units of power from participants in $S\backslash K$ to participants in $K$. In doing so, the inequality $p'_j \leq q_j' \leq p_j$ can be maintained:
\[
m' \leq \sum_{j\in K} (p_j -p'_j).
\]
The above inequality holds because when summing over $j\in S$, $p_j$'s sums to $M$ and $p'_j$'s sums to $M-p_i$.

Since $p'$ is optimal for participants $S'$ and $q'$ is not:
\begin{align}
J^\star(S') =& c_i^{(p'_i)} + \sum_{j\in S\backslash K } c_{{j}}^{(p'_j)} + \sum_{j \in K }{c_j}^{(p'_j)}\notag\\
\leq & c_i^{(p'_i)} + \sum_{j\in S\backslash K} c_j^{(p_j)} + \sum_{j\in K} c_j^{(q'_j)}=J(q'),  \label{eq:q'}
\end{align}
where we used $J(q')$ as a short-hand-notation for the cost corresponding to choosing $q'$ bids. 

 Now, we use \eqref{(i)} to replace the summations over $K$ in \eqref{eq:q'}. 
 In particular, define $q=(q_j : j\in S)$ so that 
 $$ 
 q_j := p_j +(p'_j- {q'_j})=
 \begin{cases}
 p'_j & \text{for } j \in S \backslash K,\\
 p_j + p'_j -q'_j & \text{for } j \in K.
 \end{cases}
 $$
 Note that $q$ is feasible since $p'$ and $q'$ have the same sum over $S$ (and thus cancel) and $p_j$ is feasible. 
 Further, since $(q_j - p_j)  = (p'_j- {q'_j})$, 
 by condition \eqref{(i)}: 
\begin{equation} 
\label{increas}
\sum_{j \in K}{ c_j^{(q_j)}  - c_j^{({p'_j})}} < \sum_{j \in K}{c_j^{(p_j)} - c_j^{({q'_j})}}.
\end{equation}
Adding \eqref{increas} to both side of \eqref{eq:q'} (and canceling $c_i^{(p'_i)}$) gives
\begin{align*}
J(q) &= \sum_{j\in S\backslash K} c_j^{(p'_j)} + \sum_{j\in K} c_j^{(q_j)}\\
& < \sum_{j\in S\backslash K} c_j^{(p_j)} + \sum_{j\in K} c_j^{(p_j)} = J^\star(S),
\end{align*}
which contradicts the optimality of $p$.
\qed
\end{pf}
A consequence of the above Lemma is that given the optimal allocation to procure $M$ MW, for any lower amount of MWs (while being multiple of $m$)  to be procured, the total MWs accepted from each participant never increases. Now, we are ready to prove Theorem \ref{thm:conditions_on_bids}. 

\begin{pf}
We prove that under condition \eqref{(i)}, inequality \eqref{eq:payoff_monot} holds for each $j$, for any $S \subseteq S'= \{S,i\}$ with $i$ being a generic new entered participant. Since $S$ is a generic set and $i$ can be any new participant, this is sufficient to prove that the payoffs are monotonic over all the possible couple of sets (the generalization to arbitrary sets $S'  \supseteq  S$ can then be achieved by the interpolating sums, as was done in \eqref{interpolate}). We adopt the same notation used in Lemma~\ref{lemma_regularity} and we identify with $W\subseteq S$ the set of winners.

For each $j \notin W $ we have by definition $p_{j} = 0$. 
Thus $0=J^\star(S^{-j})-J^\star(S)$, (since the optimal solution is unchanged when $j$ is removed from S). By Lemma~\ref{lemma_regularity},  ${p}'_{j}=0$ and so $0=J^\star(S')-J^\star(S')$ also. Thus, payoff monotonicity holds for $j \notin W$: $0 = J^\star(S'^{-j}) - J^\star(S') \leq J^\star(S^{-j}) - J^\star(S)= 0$ for $S'= \{S, i\} $. This says that a loser of the auction cannot become a winner as more participants enter.

For each winning participant, $w \in W$, recall that $u_{w}(S) = J^\star(S^{-w})- J^\star(S)$. Adopting the same notation of Lemma~\ref{lemma_regularity}, we can indicate it as: 
\[
 u_{w}(S):= - c_{w}^{(p_{w})}+\sum_{j \in S \setminus w} { ( c_j^{(\varrho_j)} - c_j^{(p_j)} ) } 
\]
where $\varrho_j$ are the optimal amounts to be accepted from $j \in S^{-w}$, when $w$ exits the auction. By Lemma~\ref{lemma_regularity}, in fact, $\varrho_j\geq p_j$. Similarly, after participant $i$ enters the auction, $u_{w}(S') = J^\star(S'^{-w})- J^\star(S')$. That is,
\[
 u_{w}(S'):=- c_{w}^{({p}'_{w})}  + \sum_{j \in S \setminus w} { ( c_j^{(\varrho_j')} - c_j^{({p'_j})} ) } +  c_i^{(\varrho_i')} - c_i^{({p'_i})},
\] 
 where $\varrho_i'$ are the amounts accepted from $ j \in S'^{- w}$ when $w$ exits the new auction. By Lemma~\ref{lemma_regularity}, we again have $\varrho_j'\geq {p'_j}$.
 
Notice that so far we applied Lemma~\ref{lemma_regularity} to justify the increase of the accepted amounts, first, from each $j\in S^{- w}$ and now from $j \in S'^{-w}$, due to the exit of $w$ from the auctions. We can apply Lemma~\ref{lemma_regularity} again and affirm that $ p'_j \leq p_j \hspace{1em} \forall j \in S$, and in particular $ p'_{w} \leq p_{w}$, because of the entrance of $i$.

We now find suitable lower and upper bounds to ensure  inequality  $u_{w}(S') \leq u_{w}(S)$. First, note that 
$ J^\star(S)= \sum_{ j \in S\setminus w}{ c_j^{(p_j)} } + c_{w}^{(p_{w})} \leq \sum_{ j \in S\setminus w}{ c_j^{(q_j)} } + c_{w}^{(p'_{w})}$, 
where $q_j$ are the cheapest allocation to procure $(M - p'_w)$ MW among $S\setminus w$.
By Lemma~\ref{lemma_regularity} we  have $\varrho_j \geq q_j \geq p_j \hspace{1em} \forall j \in S\setminus w$ 
, since $p_j$'s sum to $(M - p_w) \leq (M- p'_w) $ (due to $p'_w \leq p_w$), and $\varrho_j$'s sum to $M$.
Moreover, since \eqref{bids} holds and every $B_j$ satisfies \eqref{(i)}, $q_j$'s are such that $\sum_{j \in S\setminus w} ({\varrho_j}- q_j) = p'_{w}$, because exactly $M$ MW are purchased.
Using the above suboptimal allocation, we  have a lower bound for $u_{w}(S)$:  
\begin{equation}
\label{lower_bound}
u_{w}(S) \geq \sum_{j \in S \setminus w} { ( c_j^{(\varrho_j)} - c_j^{(q_j)} ) } - c_{w}^{(p'_{w})} .
\end{equation}
Defining now $\delta_j =({\varrho_j} - q_j), \forall j \in S\setminus w$ we must have $\sum_{ j\in S\setminus w} \delta_j = p'_{w}$ and $ J^\star(S'^{-w})= \sum_{j \in S \setminus w} {c_j^{(\varrho_j')} } + c_i^{(\varrho_i')} \leq   \sum_{j \in S \setminus w} {c_j^{({p'_j} + \delta_j)} } + c_i^{(p'_i)}$ , since the right hand side is a feasible cost to procure $M$ MW among the participants $\{S,i\}\setminus w$. Indeed, $\sum_{j \in S}{p'_j} + p'_i = M$ and $\sum_{j \in S \setminus w} {\delta_j} = p'_{w}$. Hence, we have: 
\begin{equation}
\label{upper_bound}
u_{w}(S') \leq \sum_{j \in S \setminus w} {(c_j^{(p'_j + \delta_j)} - c_j^{(p'_j)})} + (c_i^{(p'_i)} - c_i^{(p'_i)}) - c_{w}^{(p'_{w})} .
\end{equation}
Moreover, since $B_j$'s satisfy \eqref{(i)}, we have: 
\begin{equation} 
\label{increasing_marginal}
\forall j \in S\setminus w \hspace{1em }  ( c_j^{(p'_j + \delta_j)} - c_j^{(p'_j)} ) \leq  ( c_j^{(\varrho_j)} - c_j^{(q_j)} ).
\end{equation}
The above holds because $\forall j \in S\setminus w$ , $(\varrho_j - q_j) =( p'_j + \delta_j - p'_j) = \delta_j$ and $p'_j \leq q_j$. In particular, $p'_j$ are the amounts accepted to procure $(M-p'_{w})$ MW among $\{S,i\} \setminus w$, while $q_j$ are to procure the same MWs among $S \setminus w$. Then, combining \eqref{upper_bound} , \eqref{increasing_marginal} and \eqref{lower_bound}, we finally obtain $u_{w}(S') \leq u_{w}(S).$ \qed
\end{pf}

Condition \eqref{(i)} on every participant's bids hence is sufficient to ensure that our VCG procurement auctions will always have core outcomes. While we do not show here that the condition is necessary, we illustrate that there are certainly auctions where condition \eqref{(i)} is violated and for which payoff monotonicity does not hold. 

\begin{example}
Consider Example \ref{example_1} where power plants $PP_1$ and $PP_2$ placed just one bid for 800 MW hence violating condition  \eqref{bids}. It is easy to see that the payoffs of each of the four winners are not monotonic. In fact, if just one of them (e.g. $PP_3$) was participating, he would receive no payment; when $PP_4$,$PP_5$,$PP_6$ enter the auction, however, he becomes a winner hence making positive profit. Suppose now that $PP_1$ and $PP_2$ bid accordingly to \eqref{bids}, but the bids $B_1$ have a decreasing marginal cost: $B_1 = ( [200;400;600;800],[12'000;25'000;33'000;40'000])$, $B_2 = ([200;400;600;800],[12'000;24'000;36'000;50'000])$. In this case, when $PP_3$ participates alone, he receives a VCG payment of $40'000-33'000=7'000$ CHF; when $PP_4$,$PP_5$,$PP_6$ enter the auction, however, he receives $12'000-0=12'000$ CHF.
\end{example}

As previously anticipated, we are now able to prove that the condition derived also makes collusions and shill bidding unprofitable. Therefore, the participants are better off with  their dominant strategies, which is truthful bidding.
Although the result is well-known in literature \citep{milgrom2004putting},\citep{ausubel2006lovely} and motivates the choice of the core as a competitive standard, we can now prove it using the tools developed so far for the problem at hand. 

\begin{theorem}
\label{thm:no_collusions}
Consider a generic VCG auction $(L,w)$ cleared by \eqref{clearing}. If $ \forall j \in L $ , $B_j$ satisfies condition \eqref{(i)}. Then, 
\begin{itemize}
\item[(i)] Any group of losing bidders cannot profit by jointly lowering the bids.
\item[(ii)] Bidding with multiple identities is always unprofitable.\end{itemize}
\end{theorem}

\begin{pf}
Recall that under condition \eqref{(i)}  the participants display payoff monotonicity (Theorem \ref{thm:conditions_on_bids}).

(i) Let $C$ be a set of colluders who would lose the auction when bidding their true values $B_j =(\bar{c}_j , p_j )$, while bidding $\tilde{B}_j= ( \tilde{c_j}, p_j )$ they become winners.  Defining $B = (B_j \hspace{0.2em} j\in C,B^{-C})$ and $\tilde{B}=(\tilde{B}_j\hspace{0.2em} j\in C, B^{-C})$, the VCG payment that each player $j$ in $C$ receives is
\begin{align*} 
q_j(\tilde{B}) &= J^\star(\tilde{B}^{-j} ) - J^\star(\tilde{B}) + \tilde{c_j}^\top x_j^\star(\tilde{B} ) \\  
& \leq J^\star(B^{-C}) - J^\star(B^{-C} , \tilde{B}_j) + \tilde{c_j}^\top x_j^\star( B^{-j}, \tilde{B}_j )  \\
& =  J^\star(B^{-j}) - J^\star( B^{-j}, \tilde{B}_j ) + \tilde{c_j}^\top x_j^\star( B^{-j}, \tilde{B}_j ),
\end{align*}
where the first equality comes from definition of VCG payment, the first inequality comes from the fact that $ J^\star(\tilde{B}^{-j} ) - J^\star(\tilde{B}) \leq J^\star(B^{-C}) - J^\star(B^{-C} , \tilde{B}_j) $ since $j$ displays payoff monotonicity and $\tilde{c_j}^\top x_j^\star(\tilde{B} ) \leq \tilde{c_j}^\top x_j^\star( B^{-j}, \tilde{B}_j )$ because, when $C\setminus j$ decrease their bids, less MWs are being accepted from $j$ (Lemma \ref{lemma_regularity}) and $j$ is bidding with increasing marginal cost. The last equality comes from the fact that $C$ originally was a group of non-winners. Then, $\forall j \in C$, $q_j(\tilde{B})$ is bounded by the payment that $j$ would receive when he is the only one lowering its bid. By Theorem \ref{thm:incentive_comp}.a he will not face any benefit in doing so. 

(ii) We denote with $S \subset L$ multiple identities of the same  participant $\bar{j}$. Since every participant bids accordingly to \eqref{(i)}, the outcome is guaranteed to lie in the core. Hence, by Lemma \ref{lem:lemma_core} and substituting $u_j(B)=q_j(B)- c_j^\top x^\star(L)$, we have:
$$ \sum_{j \in S}{q_j(L)} \leq J^\star(L \setminus S) - J^\star(L) + \sum_{j \in S} {c_j^\top x_j^\star(L) },$$
where $J^\star(L \setminus S)$ is the cost when $S$, or equivalently $\bar{j}$, is removed from the auction. Therefore, the total payment that $\bar{j}$ would receive is bounded by the one he would receive bidding as a single participant. Making use of shills, hence, is not profitable. \qed
\end{pf}

To confirm the previous theoretical results, we come back to Example 1, where the TSO had to procure a fixed amount of 800 MW. That is a simple auction cleared by \eqref{eq:simpler_clearing_model}. Suppose that now power plants $PP_1$ and $PP_2$ bid according to \eqref{bids} (with $m$ = 200 MW) and condition \eqref{(i)}. The available bids are now: $B_1 = ( [200;400;600;800],[8'000;19'000;30'000;40'000])$, $B_2 = ([200;400;600;800],[12'000;24'000;36'000;50'000])$, $B_i = (200,0) \hspace{0.3em} i = 3,4,5,6$.
The winners of the auction are still power plants $PP_3$,$PP_4$,$PP_5$,$PP_6$ but now the VCG payments that they receive is $q_i(B)= 8'000 - 0 = 8'000$ CHF $i=3,4,5,6$. The total cost incurred by the TSO is much lower than before and no coalition of players now blocks the outcome. If $PP_3$,$PP_4$,$PP_5$ and $PP_6$ were multiple identities of the same losing participant (i.e. a power plant with true value greater than 40'000 CHF for 800 MW), shill bidding would become unprofitable (as expected). If, moreover, they were losing participants who jointly lowered their bids, the payments of 8'000 CHF surely made at least one of them to have negative profit.\\

\begin{figure}[h]
\begin{center}
\hspace{-0.28em}
\includegraphics[scale=0.29]{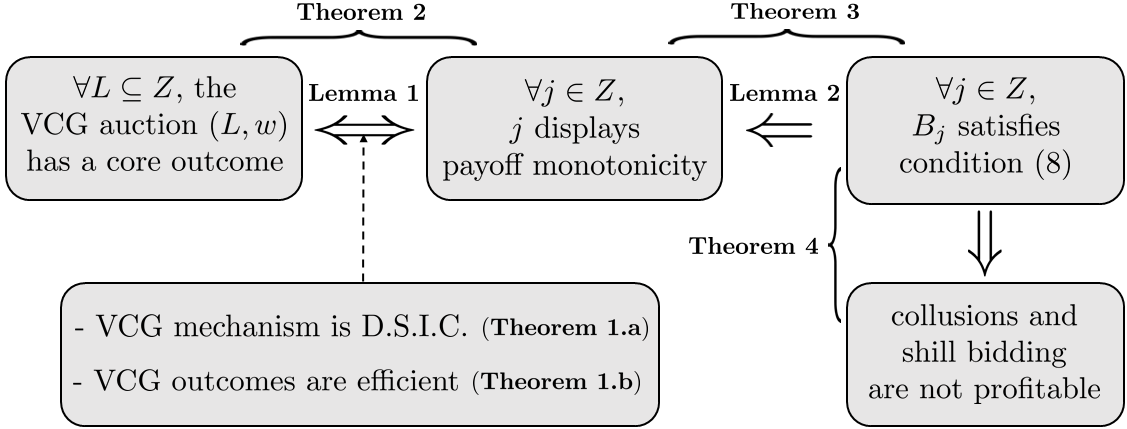}
\caption{Summarizing diagram of the relations of the results}\label{fig:diagram}
\end{center}
\end{figure}

The diagram in Fig.~\ref{fig:diagram} summarizes and links the concepts we developed so far. Notice that Lemma~\ref{lemma_regularity}, Theorem \ref{thm:conditions_on_bids} and Theorem \ref{thm:no_collusions} are specific for the class of auctions \eqref{clearing}.

\subsection{Application to two-stage stochastic market} 
As we anticipated, the Swiss reserve market as described in \citep{abbaspourtorbati2016swiss} can be modeled abstractly according to the optimization problem \eqref{eq:general_clearing_model}. There are two stages of decision variables corresponding to weekly ($x$) and daily ($y$) bids. Weekly bids are available at the instance of optimization, whereas daily bids are unknown. A number of stochastic scenarios corresponding to likely possibilities of daily bids based on their past values is used in the optimization ($\{y_i\}_{i=1}^{N_d}$). The cost function corresponds to the cost of weekly bids and the expected cost of daily bids. Thus, the cost can be written as $c^\top x + D(x,y)$. The choice function determines the accepted weekly bids. 

The function $g$ captures three types of constraints: (a) those corresponding to procurement of certain amount of tertiary reserves; (b) probabilistic constraints, which ensure that with sufficiently high probabilities, the supply and demand of power is balanced; (c) those corresponding to conditional bids. Constraint (b) links the daily and weekly variables. Constraints (a) and (c) correspond to those present in the optimization formulation \eqref{eq:simpler_clearing_model}. 

It follows from the analysis of Section \ref{sec:setup}, that the VCG mechanism applied to the two-stage stochastic market is an incentive compatible dominant strategy mechanism with socially efficient outcome.  Due to coupling of the two stage decision variables, the analysis of the core payment is significantly more difficult. In particular, the result derived in Theorem \ref{thm:conditions_on_bids} do not readily apply. The amount of procured MWs is not anymore fixed and thus \eqref{bids} is not well defined. Selecting $m$ infinitely small (forcing participants to provide continuous bid curves) and linearizing the probabilistic constraints (b), however, we could show that under condition \eqref{(i)} this clearing model follows the same regularity property of Lemma~\ref{lemma_regularity}. Whether this makes all the participants display payoff monotonicity is a subject of our current study. Nevertheless, in the numerical example section, we evaluate the performance of the VCG mechanism and compare it to the pay-as-bid mechanism. 

\section{Simulations and Analysis} 
The following simulations are based on the bids placed in the 46th Swiss weekly procurement auction of 2014, where 21 power plants bid for secondary reserves, 25 for tertiary positive and 21 for tertiary negative reserves. Note that the secondary reserves are symmetric, that is, participants need to provide same amount of positive and negative power. Tertiary reserves are on the other hand asymmetric. Thus, participants bid for tertiary negative $TRL-$, and tertiary positive $TRL+$. As in \citep{abbaspourtorbati2016swiss}, probabilistic scenarios for future daily auctions are assumed. The amount of daily reserves is based on the data of the previous week. Three scenarios are considered corresponding to nominal, high (20\% higher) and low prices (20\% lower) compared to the previous week. 

The corresponding outcome of the pay-as-bid mechanism and the VCG mechanism is shown in Table~\ref{tab:outcome_1}. Note that in reality, in a repeated bidding process, the VCG mechanism would lead to different bidding behaviors, which we have not  modeled here.  
\begin{table}[h]
\begin{center}
\caption{Outcome of the auction}\label{tab:outcome_1} 

\begin{tabular}{| l | c | c | c |}
  \hline	
  &   SRL  &  TRL- &  TRL+ \\ \hline
Procured MWs & 409 MW  & 114 MW & 100 MW \\ \hline 
\multicolumn{2}{| l |}{Sum of pay-as-bid payments} & \multicolumn{2}{| c |}{2,29 million CHF} \\ \hline
\multicolumn{2}{| l |}{Sum of VCG payments}  & \multicolumn{2}{| c |}{2,53 million CHF} \\ \hline 
\end{tabular}
\end{center}
\end{table}

Recall that in a pay-as-bid mechanism, a rational participant will overbid to ensure positive profit. Unfortunately, it is hard to know the true values of the bids for each participant.  So, it is hard to have an accurate comparison between the VCG and pay-as-bid based on past data. We now scale all the bid prices down by 90\%, assuming that those were participants' true values and hence the bids that they would have placed under the VCG mechanism. The outcome of both mechanisms is shown in Table \ref{tab:outcome_2}. 
\begin{table}[h]
\begin{center}
\caption{Outcome of the auction (scaled bids)}
\label{tab:outcome_2}

\begin{tabular}{| l | c | c | c |}
  \hline	
  &   SRL  &  TRL- &  TRL+ \\ \hline
Procured MWs & 409 MW  & 114 MW & 100 MW \\ \hline 
\multicolumn{2}{| l |}{Sum of pay-as-bid payments} & \multicolumn{2}{| c |}{2,06 million CHF} \\ \hline
\multicolumn{2}{| l |}{Sum of VCG payments}  & \multicolumn{2}{| c |}{2,28 million CHF} \\ \hline 
\end{tabular}
\end{center}
\end{table}

All the results are proportional (as it could be expected) and the sum of VCG payments is lower than the sum of pay-as-bid payments we had in the first scenario. This means that assuming such scaled bids were participants' true values, the VCG mechanism would have led to a lower procurement cost than the implemented pay-as-bid mechanism. Hence, the VCG mechanism, apart from leading to a dominant strategy equilibrium with an efficient allocation, would have been beneficial also in terms of costs, for this particular case study based on the past data. 

This does not happen in generic VCG auctions. In particular, the cost incurred by the auctioneer in a VCG auction is usually higher than the cost under a pay-as-bid mechanism, considering the same set of bids. To see this, recall that the VCG payments are $q_j(B) = [ J^\star(B^{-j}) - J^\star(B)] + c_j^\top x_j^\star(B)$. These payments  measure the benefit that each participant brings to the auction. When the VCG payments are computed through the  two-stage stochastic optimization algorithm of \citep{abbaspourtorbati2016swiss} we observed that the costs are not significantly different from the pay-as-bid mechanism. Intuitively, the two-stage market  softens the benefit that every participant brings to the weekly auction: his accepted bids can always be replaced by amounts of MWs allocated to the future. In fact, the amounts of MWs bought in the weekly auction are not fixed and they are flexible depending on the future daily bids available in that week.

To confirm the intuition above, we now assume that we had \emph{perfect information} about the future daily auctions. As such, we run a deterministic auction assuming that the TSO already knew that the optimal amounts to be purchased were 409 MW for SRL , 114 MW for TRL- and 100 MW for TRL+ as predicted in Table \ref{tab:outcome_1} . Given fixed MWs to be procured, the auction is cleared by the simplified model \eqref{eq:simpler_clearing_model}. In this case, naturally, we have the same winners of the auction as the previous case for both VCG and pay-as-bid mechanism. The VCG payments however are significantly higher than the pay-as-bid payments. The results are shown in Table \ref{tab:outcome_3}.  

\begin{table}[h]
\begin{center}
\label{tab:outcome_3}
\caption{Outcome of the deterministic auction}
\begin{tabular}{| l | c | c | c |}
  \hline	
  &   SRL  &  TRL- &  TRL+ \\ \hline
Procured MWs & 409 MW  & 114 MW & 100 MW \\ \hline 
\multicolumn{2}{| l |}{Sum of pay-as-bid payments} & \multicolumn{2}{| c |}{2,06 million CHF} \\ \hline
\multicolumn{2}{| l |}{Sum of VCG payments}  & \multicolumn{2}{| c |}{7,97 million CHF} \\ \hline 
\end{tabular}
\end{center}
\end{table}
The result can be explained as follows. When a winner $j$ is removed from the auction (to compute the term $J^\star(B^{-j})$) the amounts of MWs to be accepted among the other participants originally were subject to flexibility due to two stage decision variables and lack of a fixed total amount for each type of reserve SRL, TRL-, TRL+.  If these total reserves are fixed for each type,  the benefit that every participant brings to the Swiss weekly auction is much higher, and this results  in higher VCG payments. 

The mixed integer optimization problems  were solved with GUROBI, on a quad-core computer with processing speed 1.7 GHz and memory 4 Gb. The first two simulations had a computation time of 9 min, with an average of 17 s for each optimal cost $J^\star$. The last simulation took 7 min, with an average of 14 s for each $J^\star$. 

\section{Conclusion}
We developed a VCG market mechanism for electricity markets, motivated by the set-up of the control reserves (ancillary services) market. We showed that the mechanism results in an incentive compatible dominant strategy Nash equilibrium. Furthermore, this mechanism is socially efficient. Through examples, we showed that shill bidding can occur. We thus, derived conditions under which a deterministic procurement mechanism can guarantee no shill bidding and thus competitive outcomes. These findings, both theoretical and empirical, act to support the application of VCG mechanism for the electricity markets under consideration. By removing incentives for collusion and by providing a simple truthful mechanism, we expect that the implementation simplifies the biding process, increases markets efficiency and encourages participation from increasing number of entities. 
We verified our results based on market data available from Swissgrid. Future work consists of deriving conditions under which Theorem \ref{thm:no_collusions} could be extended to the stochastic market. 

\begin{ack}
We thank Farzaneh Abbaspour and Marek Zima from Swissgrid for helpful discussions. 
\vspace{2em}
\end{ack}

\bibliography{ref_vcg_power}             
                                                   







\end{document}